\documentclass[%
 twocolumn,
 superscriptaddress,
 amsmath,amssymb,
 prb,
]{revtex4-1}

\usepackage[dvips]{graphicx} % Include figure files
\usepackage{bm}% bold math
\usepackage{float}

\graphicspath{{./}{Figures//}{Figures/}}
\DeclareGraphicsExtensions{.eps}

\newcommand {\be}{\begin{eqnarray}}
\newcommand {\ee}{\end{eqnarray}}

\begin{document}

%\preprint{APS/123-QED}

\title{Comparison of charge modulations in La$_{1.875}$Ba$_{0.125}$CuO$_4$ and YBa$_2$Cu$_3$O$_{6.6}$}

\author{V. Thampy}
\affiliation{Condensed Matter Physics and Materials Science Department, Brookhaven National Laboratory, Upton, New York 11973, USA}

\author{S. Blanco-Canosa}
\affiliation{Max-Planck-Institut f\"ur Festk\"orperforschung, Heisenbergstrasse 1, D-70569 Stuttgart, Germany}

\author{M. Garc\'ia-Fern\'andez}
\author{M. P. M. Dean}
\author{G. D. Gu}
\affiliation{Condensed Matter Physics and Materials Science Department, Brookhaven National Laboratory, Upton, New York 11973, USA}

\author{M. F\"{o}erst}
\affiliation{Max-Planck Institute for the Structure and Dynamics of Matter, Hamburg, Germany}

\author{B. Keimer}
\author{M. Le Tacon}
\affiliation{Max-Planck-Institut f\"ur Festk\"orperforschung, Heisenbergstrasse 1, D-70569 Stuttgart, Germany}

\author{S. B. Wilkins}
\author{J. P. Hill}
\affiliation{Condensed Matter Physics and Materials Science Department, Brookhaven National Laboratory, Upton, New York 11973, USA}

\date{\today}

\begin{abstract}
A charge modulation has recently been reported in (Y,Nd)Ba$_2$Cu$_3$O$_{6+x}$ [Ghiringhelli {\em et al.} Science 337, 821 (2013)]. Here we report Cu $L_3$ edge soft x-ray scattering studies comparing the lattice modulation associated with the charge modulation in YBa$_2$Cu$_3$O$_{6.6}$ with that associated with the well known charge and spin stripe order in La$_{1.875}$Ba$_{0.125}$CuO$_4$. We find that the correlation length in the CuO$_2$ plane is isotropic in both cases, and is $259 \pm 9$~\AA~for La$_{1.875}$Ba$_{0.125}$CuO$_4$ and $55 \pm 15$~\AA~for YBa$_2$Cu$_3$O$_{6.6}$. Assuming weak inter-planar correlations of the charge ordering in both compounds, we conclude that the order parameters of the lattice modulations in La$_{1.875}$Ba$_{0.125}$CuO$_4$ and YBa$_2$Cu$_3$O$_{6.6}$ are of the same order of magnitude. 
\end{abstract}

\maketitle

Doped cuprates in the so-called 214 family, i.e.\ La$_2$CuO$_4$ with dopants substituting La (La$_{2-x-y}$(Sr,Ba)$_x$(Eu,Nd)$_y$CuO$_4$), have been shown to exhibit so-called stripe order near $x=1/8$: that is anti-phase antiferromagnetic domains separated by stripes of uniaxial charge [\onlinecite{Tranquada1995, PhysRevB.70.104517, PhysRevB.78.174529, Abbamonte2005, PhysRevB.77.064520, PhysRevB.83.104506, Tranquada1995, PhysRevB.79.100502, PhysRevB.83.092503}]. This ordering co-exists with, and possibly competes with superconductivity [\onlinecite{Tranquada1995, PhysRevLett.99.067001, PhysRevB.78.174529, KOJIMA, PhysRevB.83.104506}]. The stripes of LBCO break rotational symmetry and thus also bear some similarity to the nematic and smectic order observed in the underdoped pseudogap region [\onlinecite{Daou2010, Lawler2010}]. Further, the recent observation of charge density wave (CDW) correlations in the 123 family ((Y,Nd)Ba$_2$Cu$_3$O$_{6+x}$) [\onlinecite{Ghiringhelli17082012, Chang2012a}] of the cuprates suggests that charge ordering may in fact be a universal feature of cuprate superconductors. However, it is not clear how the charge modulations in the 123 compounds relate to the more familiar stripes observed in the 214 family.

In particular, the modulations observed in La$_{1.875}$Ba$_{0.125}$CuO$_4$ (LBCO) and YBa$_2$Cu$_3$O$_{6.6}$ (YBCO) differ in some important respects. First, the characteristic in-plane wave-vectors are different: in LBCO, which is tetragonal, the modulation peaks at the same position, $q \approx 0.24$, along both $H$ and $K$ [\onlinecite{Tranquada1995, PhysRevB.83.104506, PhysRevB.84.195101}]. In YBCO, which is orthorhombic, however, $q$ is closer to 0.31. Further, there is a small anisotropy in the value of q between $H$ and $K$ in r.l.u., though it has the same value in~\AA$^{-1}$ [\onlinecite{Chang2012a, Achkar2012, PhysRevLett.110.137004, Ghiringhelli17082012, PhysRevLett.110.187001}]. Second, in LBCO, the appearance of stripe ordering coincides with the onset of a low temperature tetragonal phase (LTT) [\onlinecite{Tranquada1995, PhysRevB.83.104506}]. The stripe order, if it exists, is very weak above the LTT transition temperature ($T_{LTT} = 68 $K) [\onlinecite{PhysRevB.83.104506}]. Below the transition temperature, the charge stripe order parameter is constant as a function of temperature [\onlinecite{PhysRevB.84.195101}]. In YBCO, there is no such concomitant structural transition [\onlinecite{Ghiringhelli17082012}], and the amplitude and correlation length of the charge order peak increase with decreasing temperature, reaching a maximum at the superconducting transition temperature, below which they both decrease [\onlinecite{Ghiringhelli17082012}]. Finally, the doping dependence of the in-plane wave-vectors in the two systems is quite different. In LBCO there is a strong positive correlation between the doping content ($x$) and $q$, particularly below $x = 1/8$ [\onlinecite{PhysRevB.77.224410, PhysRevB.83.104506}]. In YBCO, the dependence on doping, though weaker, is the opposite, i.e.\ $q$ decreases with increased doping [\onlinecite{PhysRevLett.110.137004, PhysRevLett.110.187001}]. 

Given these distinctions, a rigorous comparison of the ordering between the two compounds is required to shed light on whether or not a common underlying instability gives rise to these charge correlations. Here, we report such a comparison using resonant x-ray scattering (RXS) at the Cu $L_3$ edge, which is especially effective for detecting charge order and/or corresponding lattice modulations [\onlinecite{PhysRevB.74.195113}].  The two samples (LBCO and YBCO), well characterized by other measurements, are studied with the same experimental set up and under identical conditions to allow direct, quantitative, comparison of the scattered intensities. We find that the in-plane correlation lengths are isotropic in both cases, $\xi^{\mathrm{In-plane}}_{\mathrm{LBCO}} = 259 \pm 9$~\AA, $\xi^{\mathrm{In-plane}}_{\mathrm{YBCO}} = 55 \pm 15$~\AA, and that the order parameters are of the same order of magnitude in the two systems.   

The LBCO single crystal used for this experiment was grown using the floating zone method at Brookhaven National Lab, USA. Its crystal structure is tetragonal with space group ($I4/mmm$) and lattice parameters $a = b = 3.78$~\AA, $c = 13.28$~\AA~at room temperature. Throughout this paper LBCO is indexed with this unit cell. The YBCO single crystal, which is the same sample as in a previous study [\onlinecite{Ghiringhelli17082012}], is detwinned and was synthesized by a self-flux method at the Max Planck Institute, Stuttgart [\onlinecite{Hinkov2007}]. Its crystal structure is orthorhombic with lattice parameters $a = 3.82$~\AA, $b = 3.88$~\AA, $c = 11.7$~\AA. Both the LBCO and the YBCO samples have the same hole concentration ($p \approx 0.125$). In LBCO, this hole concentration corresponds to the dip in the dome of superconductivity in the phase diagram where the stripe order is strongest [\onlinecite{PhysRevB.83.104506}], while in YBCO, this corresponds to the point where there is a plateau in the phase diagram [\onlinecite{PhysRevB.73.180505}]. 

\begin{figure}[b]
    \includegraphics*[trim = 0mm 0mm 0mm 0mm, clip,width=.99\columnwidth]{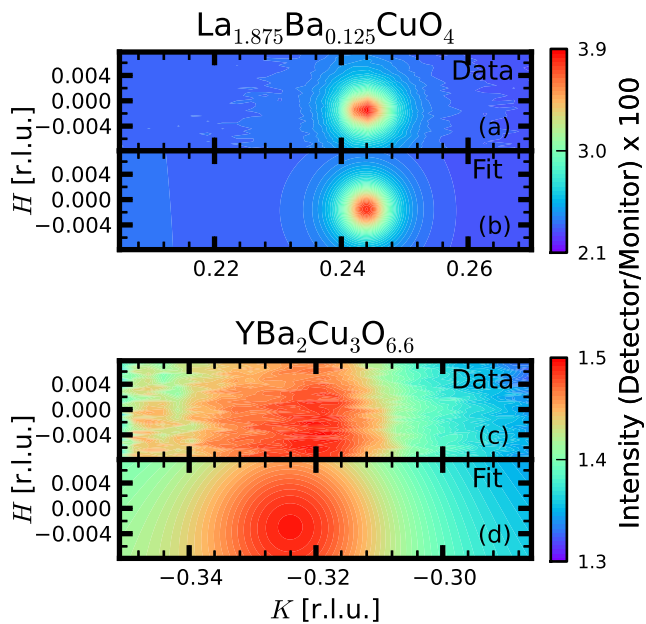}
	\caption{(a) 2D map of the scattering intensity for LBCO in the $(H, K, 1.39)_{HTT}$ plane integrated over the region $1.35 \le L \le 1.43$, showing isotropic correlation in the $a-b$ plane. Data were taken at $T$ = 15 K. (b) Result of a 2D fit of (a) to a Lorentzian squared function with a planar background. (c) and (d) show the scattering and fit for YBCO. The data for YBCO were taken at $T$ = 60 K. Data for both LBCO and YBCO were collected at the energy corresponding to the peak in the absorption spectrum.} 
\label{fig:slices}
\end{figure}

Soft x-ray diffraction experiments were carried out on the X1A2 beamline at the NSLS, Brookhaven National Laboratory, USA, using a six-circle in-vacuum diffractometer. The YBCO single crystal was polished in air [\onlinecite{Ghiringhelli17082012}], while the LBCO crystal was cleaved {\em ex situ}, to reveal surfaces  with a [001] surface normal. They were mounted such that the [001] and [010] directions lay in the scattering plane. Since the YBCO sample is well detwinned, the [010] direction was chosen to avoid the strong background from the oxygen chain superstructures along the [100] direction [\onlinecite{PhysRevLett.110.187001, Ghiringhelli17082012}]. Experiments were performed in a vertical scattering geometry with $\sigma-$incident x rays, that is, the electric field of the incident x-ray photons was always along the [100] direction ($a$ axis), within the CuO$_2$ planes and perpendicular to the scattering plane. The samples were cooled in a He flow cryostat, and the scattered x-rays were detected using an in vacuum CCD camera at a fixed distance of 0.355~m from the sample. The CCD camera has (2048 x 2048) pixels, each pixel is ($13.5 \times 13.5) \mu m^2$ in size.  

\begin{figure}[b]
    \includegraphics*[trim = 0mm 0mm 0mm 0mm, clip,width=0.99\columnwidth]{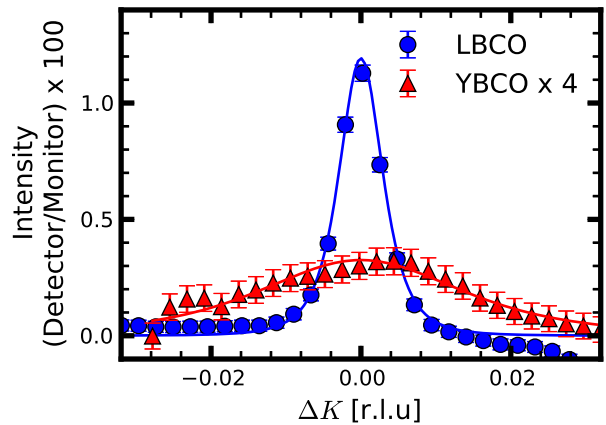}
	\caption{Line cuts of the background subtracted scattered intensities for LBCO (blue) and YBCO (red). The YBCO data has been multiplied by a factor of 4 for clarity. The x-axis shows the displacement in r.l.u. from the center of the peaks. The solid lines show the line cuts of the 2D Lorentzian squared fit which was shown in Fig.~\ref{fig:slices}(b) and Fig.~\ref{fig:slices}(d). The error bars shown here do not show statistical errors, but an estimate of the systematic errors of the experiment.}    
    \label{fig:cuts}
\end{figure}

The diffraction data were collected using photons with energies near the Cu $L_3$ absorption edge ($2p_{3/2} \rightarrow 3d$) which, as noted earlier, greatly enhances the sensitivity of the scattering signal to the lattice distortion induced by the charge ordering [\onlinecite{PhysRevLett.61.1245}]. More specifically, the resonant signal from each material was maximized by choosing the energy corresponding to the peak intensity in the x-ray absorption spectrum (XAS). For LBCO, the data were collected at 15~K, while for YBCO, the data were collected at 60~K, the temperature at which the peak scattered intensity from the charge modulation is known to be largest [\onlinecite{Ghiringhelli17082012}]. The beam-line configuration used was identical for both samples, and the data were normalized to an incident intensity beam monitor. The sample orientation (UB matrix) was determined using the (002) and (101) Bragg reflections measured at 1060~eV and 1700~eV respectively. The momentum dependent scattering for both LBCO and YBCO was measured using $K$ scans, keeping $H$ and $L$ fixed. The positive $K$ values at which the data is reported for LBCO correspond to grazing exit geometry, while the negative $K$ values for YBCO correspond to grazing incidence geometry. The same value of $L$ was chosen for both samples ($L$ = 1.39 r.l.u.). The CCD image collected at each point was then converted into reciprocal lattice coordinates pixel by pixel, from which two dimensional slices through reciprocal space were constructed. These are shown in Figs.~\ref{fig:slices}[a] and ~\ref{fig:slices}[c] for LBCO and YBCO respectively. We note that the scattered intensities shown here are not energy resolved, and have some contribution from inelastic scattering as well. However, the inelastic scattering comes mainly from inter-orbital transitions ($dd$ excitations), with small contributions from paramagnon and charge transfer excitations  [\onlinecite{NPh000294485400021, 2013arXiv1305.1317D, PhysRevLett.92.117406, Ghiringhelli17082012}]. The inelastic contribution is only weakly dependent on $q$ and can be subtracted as a flat background.  

For a more accurate comparison of total scattering cross section from the two samples, two corrections were applied to the measured scattering intensity, in addition to the flat background subtraction. First, the data were corrected for self-absorption using: $ I = I_o(1 + \frac{\mathrm{sin}~\theta_i}{\mathrm{sin}~\theta_f})$, where $I$ is the corrected intensity, $I_o$ is the measured intensity, $\theta_i$ is the angle between the sample surface and the incident wave-vector $\bm{k}_i$, and $\theta_f$ is the angle between the sample surface and the scattered wave-vector $\bm{k}_f$ [\onlinecite{2013arXiv1303.5359D}]. Due to the scattering geometry and the different wave-vectors probed for LBCO and YBCO, this factor was larger for LBCO than YBCO by a factor of $\sim 2.5$. Also, a multiplicative correction factor of $\sim 0.6$, was applied to the YBCO intensity to account for the different absorption coefficients for the two samples, and consequent different volumes probed. This factor was calculated from tabulated values of the attenuation lengths for the two compounds. 

Fig.~\ref{fig:slices}[a] shows the momentum dependence of the scattering in the HK plane for LBCO, integrated over the region $1.35 \le L \le 1.43$. Note that all intensities hence forth are integrated over the same range in $L$, unless otherwise noted. The scale bars for Fig.~\ref{fig:slices} show the total scattering intensity normalized to the monitor and may be compared directly. These data were taken at $T = 15$~K. Fig.~\ref{fig:slices}[c] shows the scattering for YBCO which was measured at $T = 60$~K. As noted earlier, in both cases the energy at which the data were collected corresponds to the peak of the XAS for the respective materials. It is immediately clear that the peak in YBCO is much broader than in LBCO, signaling a much shorter correlation length of the modulation in the 123 compound. Further, the peak intensity of the scattering for LBCO is over an order of magnitude larger than that for YBCO.

For a more quantitative comparison, these data were fit to a 2D Lorentzian squared function with a planar background. Figs.~\ref{fig:slices}[b]([d]) show the corresponding fits for LBCO(YBCO). In the fits shown here, the widths along $H$ and $K$ are constrained to be identical. When the widths of the peak along $H$ and $K$ were allowed to vary independently, the goodness of the fit as measured by $\chi^2$, improves by only 2~\% and the resulting fit widths were equal to within 10~\%. Therefore, to reduce the number of free parameters, we constrain the widths to be equal, that is, the correlation lengths are isotropic in the Cu-O plane. 

The fits yield a peak intensity $\sim$17 times larger for LBCO than for YBCO. For LBCO, the peak is centered at $K$ = 0.244(1) r.l.u., while for YBCO the center is at $K$ = -0.323(2) r.l.u.. Also, we extract in-plane correlation lengths ($\frac{1}{\mathrm{HWHM}}$) of $259 \pm 9$~\AA~for LBCO and $54.7 \pm 15$~\AA~for YBCO. That is, the correlations in YBCO, extending to $\sim 15$ unit cells, are about 5 times shorter in range than in LBCO, where they extend up to $\sim 70$ unit cells. These results are consistent with earlier reports in LBCO [\onlinecite{PhysRevB.84.195101}] and YBCO [\onlinecite{Ghiringhelli17082012}]. We note here that the correlation length ($255 \pm 5$~\AA) for LBCO reported in [\onlinecite{PhysRevB.84.195101}] was measured at the oxygen $K$-edge prepeak, which directly probes the charge modulation. The comparison between the scattered intensity peaks and widths for the two samples is seen more clearly in Fig.~\ref{fig:cuts} which shows line cuts through the peaks for LBCO and YBCO with the planar background subtracted. 

We now discuss the question of the relative intensities of the two modulations, which relates to the size of the square of the two order parameters. Without having a precise form of the structural distortion associated with the charge modulation, it is difficult to directly relate the scattering intensity to the modulation amplitude. Here we make the simplifying assumption that at the copper edge, the structure factors have the same form for both LBCO and YBCO. Further, we make the assumption that the volume fraction of the charge modulation is the same in both materials. Then, the two order parameters are proportional to the integrated intensity per unit volume probed and we can compare the two directly. The correlations along the $c$ axis being weak [\onlinecite{PhysRevB.84.195101}], we can further assume that the $L$-axis dependence of the scattering is dominated by the structure factor, which we assume to be the same for the two materials . Since the range of $L$ over which the intensity is integrated is the same for both LBCO and YBCO, the $L$ dependence can be factored out in the comparison. Therefore, the integrated intensity is $\propto I_{\mathrm{peak}} \times \Gamma^2$, where $\Gamma$ is the in-plane peak width and $I_{peak}$ is the peak intensity. We then find $\frac{I_{\mathrm{YBCO}}}{I_{\mathrm{LBCO}}} = 1.3 \pm 0.5$. Or equivalently, $\frac{A_{\mathrm{YBCO}}}{A_{\mathrm{LBCO}}} = 1.1 \pm 0.5$, where $A$ is the amplitude of the respective order parameters, that is the amplitude of the associated lattice modulation. Thus, even though the peak intensity of the modulation is much weaker in YBCO, as are the correlations, the order parameters in the two materials are approximately equal. 

Unfortunately, from these measurements alone it is not possible to determine the amplitude of the respective charge modulations driving these lattice modulations. This is an interesting question and there are perhaps methods to answer this empirically [\onlinecite{PhysRevB.74.195113}], or possibly through calculations of the momentum-dependent electron-phonon coupling. However, we leave this for future studies.
  
A previous study [\onlinecite{PhysRevB.84.195101}] highlighted the huge difference in the amplitude of the charge modulations associated with stripe order between two 214 systems, LBCO and LNSCO. Here, we have two systems, in which as noted earlier, the charge correlations differ in significant ways, the most important being, perhaps, the different wave-vectors for the modulations. Another important distinction is the absence of magnetic order in YBCO for the doping considered here [\onlinecite{PhysRevB.81.104507, 0034-4885-61-10-002, 1367-2630-12-10-105006}], as opposed to LBCO where the uniaxial arrangement of the charges delineates the anti-ferromagnetically ordered regions to form the so-called stripes. Despite these distinctions, we find that the amplitudes of the associated lattice modulations in the two systems are comparable in magnitude. More work is required in order to determine whether this fact is a simple coincidence or whether this is hinting towards some similarities in the mechanism for the formation of charge order in LBCO and YBCO.

Work performed at BNL was supported by the US Department of Energy, Division of Materials Science, under Contract No. DE- AC02-98CH10886.

\bibliography{YBCO_v_LBCO}

\begin{thebibliography}{30}%
\makeatletter
\providecommand \@ifxundefined [1]{%
 \@ifx{#1\undefined}
}%
\providecommand \@ifnum [1]{%
 \ifnum #1\expandafter \@firstoftwo
 \else \expandafter \@secondoftwo
 \fi
}%
\providecommand \@ifx [1]{%
 \ifx #1\expandafter \@firstoftwo
 \else \expandafter \@secondoftwo
 \fi
}%
\providecommand \natexlab [1]{#1}%
\providecommand \enquote  [1]{``#1''}%
\providecommand \bibnamefont  [1]{#1}%
\providecommand \bibfnamefont [1]{#1}%
\providecommand \citenamefont [1]{#1}%
\providecommand \href@noop [0]{\@secondoftwo}%
\providecommand \href [0]{\begingroup \@sanitize@url \@href}%
\providecommand \@href[1]{\@@startlink{#1}\@@href}%
\providecommand \@@href[1]{\endgroup#1\@@endlink}%
\providecommand \@sanitize@url [0]{\catcode `\\12\catcode `\$12\catcode
  `\&12\catcode `\#12\catcode `\^12\catcode `\_12\catcode `\%12\relax}%
\providecommand \@@startlink[1]{}%
\providecommand \@@endlink[0]{}%
\providecommand \url  [0]{\begingroup\@sanitize@url \@url }%
\providecommand \@url [1]{\endgroup\@href {#1}{\urlprefix }}%
\providecommand \urlprefix  [0]{URL }%
\providecommand \Eprint [0]{\href }%
\providecommand \doibase [0]{http://dx.doi.org/}%
\providecommand \selectlanguage [0]{\@gobble}%
\providecommand \bibinfo  [0]{\@secondoftwo}%
\providecommand \bibfield  [0]{\@secondoftwo}%
\providecommand \translation [1]{[#1]}%
\providecommand \BibitemOpen [0]{}%
\providecommand \bibitemStop [0]{}%
\providecommand \bibitemNoStop [0]{.\EOS\space}%
\providecommand \EOS [0]{\spacefactor3000\relax}%
\providecommand \BibitemShut  [1]{\csname bibitem#1\endcsname}%
\let\auto@bib@innerbib\@empty
%</preamble>
\bibitem [{\citenamefont {Tranquada}\ \emph {et~al.}(1995)\citenamefont
  {Tranquada}, \citenamefont {Sternlieb},\ and\ \citenamefont
  {Axe}}]{Tranquada1995}%
  \BibitemOpen
  \bibfield  {author} {\bibinfo {author} {\bibfnamefont {J.}~\bibnamefont
  {Tranquada}}, \bibinfo {author} {\bibfnamefont {B.}~\bibnamefont
  {Sternlieb}}, \ and\ \bibinfo {author} {\bibfnamefont {J.}~\bibnamefont
  {Axe}},\ }\href
  {http://www.nature.com/nature/journal/v375/n6532/abs/375561a0.html}
  {\bibfield  {journal} {\bibinfo  {journal} {Nature}\ }\textbf {\bibinfo
  {volume} {375}},\ \bibinfo {pages} {561} (\bibinfo {year}
  {1995})}\BibitemShut {NoStop}%
\bibitem [{\citenamefont {Fujita}\ \emph {et~al.}(2004)\citenamefont {Fujita},
  \citenamefont {Goka}, \citenamefont {Yamada}, \citenamefont {Tranquada},\
  and\ \citenamefont {Regnault}}]{PhysRevB.70.104517}%
  \BibitemOpen
  \bibfield  {author} {\bibinfo {author} {\bibfnamefont {M.}~\bibnamefont
  {Fujita}}, \bibinfo {author} {\bibfnamefont {H.}~\bibnamefont {Goka}},
  \bibinfo {author} {\bibfnamefont {K.}~\bibnamefont {Yamada}}, \bibinfo
  {author} {\bibfnamefont {J.~M.}\ \bibnamefont {Tranquada}}, \ and\ \bibinfo
  {author} {\bibfnamefont {L.~P.}\ \bibnamefont {Regnault}},\ }\href {\doibase
  10.1103/PhysRevB.70.104517} {\bibfield  {journal} {\bibinfo  {journal} {Phys.
  Rev. B}\ }\textbf {\bibinfo {volume} {70}},\ \bibinfo {pages} {104517}
  (\bibinfo {year} {2004})}\BibitemShut {NoStop}%
\bibitem [{\citenamefont {Tranquada}\ \emph {et~al.}(2008)\citenamefont
  {Tranquada}, \citenamefont {Gu}, \citenamefont {H\"ucker}, \citenamefont
  {Jie}, \citenamefont {Kang}, \citenamefont {Klingeler}, \citenamefont {Li},
  \citenamefont {Tristan}, \citenamefont {Wen}, \citenamefont {Xu},
  \citenamefont {Xu}, \citenamefont {Zhou},\ and\ \citenamefont
  {v.~Zimmermann}}]{PhysRevB.78.174529}%
  \BibitemOpen
  \bibfield  {author} {\bibinfo {author} {\bibfnamefont {J.~M.}\ \bibnamefont
  {Tranquada}}, \bibinfo {author} {\bibfnamefont {G.~D.}\ \bibnamefont {Gu}},
  \bibinfo {author} {\bibfnamefont {M.}~\bibnamefont {H\"ucker}}, \bibinfo
  {author} {\bibfnamefont {Q.}~\bibnamefont {Jie}}, \bibinfo {author}
  {\bibfnamefont {H.-J.}\ \bibnamefont {Kang}}, \bibinfo {author}
  {\bibfnamefont {R.}~\bibnamefont {Klingeler}}, \bibinfo {author}
  {\bibfnamefont {Q.}~\bibnamefont {Li}}, \bibinfo {author} {\bibfnamefont
  {N.}~\bibnamefont {Tristan}}, \bibinfo {author} {\bibfnamefont {J.~S.}\
  \bibnamefont {Wen}}, \bibinfo {author} {\bibfnamefont {G.~Y.}\ \bibnamefont
  {Xu}}, \bibinfo {author} {\bibfnamefont {Z.~J.}\ \bibnamefont {Xu}}, \bibinfo
  {author} {\bibfnamefont {J.}~\bibnamefont {Zhou}}, \ and\ \bibinfo {author}
  {\bibfnamefont {M.}~\bibnamefont {v.~Zimmermann}},\ }\href {\doibase
  10.1103/PhysRevB.78.174529} {\bibfield  {journal} {\bibinfo  {journal} {Phys.
  Rev. B}\ }\textbf {\bibinfo {volume} {78}},\ \bibinfo {pages} {174529}
  (\bibinfo {year} {2008})}\BibitemShut {NoStop}%
\bibitem [{\citenamefont {Abbamonte}\ \emph {et~al.}(2005)\citenamefont
  {Abbamonte}, \citenamefont {Rusydi}, \citenamefont {Smadici}, \citenamefont
  {Gu}, \citenamefont {Sawatzky},\ and\ \citenamefont {Feng}}]{Abbamonte2005}%
  \BibitemOpen
  \bibfield  {author} {\bibinfo {author} {\bibfnamefont {P.}~\bibnamefont
  {Abbamonte}}, \bibinfo {author} {\bibfnamefont {A.}~\bibnamefont {Rusydi}},
  \bibinfo {author} {\bibfnamefont {S.}~\bibnamefont {Smadici}}, \bibinfo
  {author} {\bibfnamefont {G.~D.}\ \bibnamefont {Gu}}, \bibinfo {author}
  {\bibfnamefont {G.~A.}\ \bibnamefont {Sawatzky}}, \ and\ \bibinfo {author}
  {\bibfnamefont {D.~L.}\ \bibnamefont {Feng}},\ }\href {\doibase
  10.1038/nphys178} {\bibfield  {journal} {\bibinfo  {journal} {Nature
  Physics}\ }\textbf {\bibinfo {volume} {1}},\ \bibinfo {pages} {155} (\bibinfo
  {year} {2005})}\BibitemShut {NoStop}%
\bibitem [{\citenamefont {Kim}\ \emph {et~al.}(2008)\citenamefont {Kim},
  \citenamefont {Gu}, \citenamefont {Gog},\ and\ \citenamefont
  {Casa}}]{PhysRevB.77.064520}%
  \BibitemOpen
  \bibfield  {author} {\bibinfo {author} {\bibfnamefont {Y.-J.}\ \bibnamefont
  {Kim}}, \bibinfo {author} {\bibfnamefont {G.~D.}\ \bibnamefont {Gu}},
  \bibinfo {author} {\bibfnamefont {T.}~\bibnamefont {Gog}}, \ and\ \bibinfo
  {author} {\bibfnamefont {D.}~\bibnamefont {Casa}},\ }\href {\doibase
  10.1103/PhysRevB.77.064520} {\bibfield  {journal} {\bibinfo  {journal} {Phys.
  Rev. B}\ }\textbf {\bibinfo {volume} {77}},\ \bibinfo {pages} {064520}
  (\bibinfo {year} {2008})}\BibitemShut {NoStop}%
\bibitem [{\citenamefont {H\"ucker}\ \emph {et~al.}(2011)\citenamefont
  {H\"ucker}, \citenamefont {v.~Zimmermann}, \citenamefont {Gu}, \citenamefont
  {Xu}, \citenamefont {Wen}, \citenamefont {Xu}, \citenamefont {Kang},
  \citenamefont {Zheludev},\ and\ \citenamefont
  {Tranquada}}]{PhysRevB.83.104506}%
  \BibitemOpen
  \bibfield  {author} {\bibinfo {author} {\bibfnamefont {M.}~\bibnamefont
  {H\"ucker}}, \bibinfo {author} {\bibfnamefont {M.}~\bibnamefont
  {v.~Zimmermann}}, \bibinfo {author} {\bibfnamefont {G.~D.}\ \bibnamefont
  {Gu}}, \bibinfo {author} {\bibfnamefont {Z.~J.}\ \bibnamefont {Xu}}, \bibinfo
  {author} {\bibfnamefont {J.~S.}\ \bibnamefont {Wen}}, \bibinfo {author}
  {\bibfnamefont {G.}~\bibnamefont {Xu}}, \bibinfo {author} {\bibfnamefont
  {H.~J.}\ \bibnamefont {Kang}}, \bibinfo {author} {\bibfnamefont
  {A.}~\bibnamefont {Zheludev}}, \ and\ \bibinfo {author} {\bibfnamefont
  {J.~M.}\ \bibnamefont {Tranquada}},\ }\href {\doibase
  10.1103/PhysRevB.83.104506} {\bibfield  {journal} {\bibinfo  {journal} {Phys.
  Rev. B}\ }\textbf {\bibinfo {volume} {83}},\ \bibinfo {pages} {104506}
  (\bibinfo {year} {2011})}\BibitemShut {NoStop}%
\bibitem [{\citenamefont {Fink}\ \emph {et~al.}(2009)\citenamefont {Fink},
  \citenamefont {Schierle}, \citenamefont {Weschke}, \citenamefont {Geck},
  \citenamefont {Hawthorn}, \citenamefont {Soltwisch}, \citenamefont {Wadati},
  \citenamefont {Wu}, \citenamefont {D\"urr}, \citenamefont {Wizent},
  \citenamefont {B\"uchner},\ and\ \citenamefont
  {Sawatzky}}]{PhysRevB.79.100502}%
  \BibitemOpen
  \bibfield  {author} {\bibinfo {author} {\bibfnamefont {J.}~\bibnamefont
  {Fink}}, \bibinfo {author} {\bibfnamefont {E.}~\bibnamefont {Schierle}},
  \bibinfo {author} {\bibfnamefont {E.}~\bibnamefont {Weschke}}, \bibinfo
  {author} {\bibfnamefont {J.}~\bibnamefont {Geck}}, \bibinfo {author}
  {\bibfnamefont {D.}~\bibnamefont {Hawthorn}}, \bibinfo {author}
  {\bibfnamefont {V.}~\bibnamefont {Soltwisch}}, \bibinfo {author}
  {\bibfnamefont {H.}~\bibnamefont {Wadati}}, \bibinfo {author} {\bibfnamefont
  {H.-H.}\ \bibnamefont {Wu}}, \bibinfo {author} {\bibfnamefont {H.~A.}\
  \bibnamefont {D\"urr}}, \bibinfo {author} {\bibfnamefont {N.}~\bibnamefont
  {Wizent}}, \bibinfo {author} {\bibfnamefont {B.}~\bibnamefont {B\"uchner}}, \
  and\ \bibinfo {author} {\bibfnamefont {G.~A.}\ \bibnamefont {Sawatzky}},\
  }\href {\doibase 10.1103/PhysRevB.79.100502} {\bibfield  {journal} {\bibinfo
  {journal} {Phys. Rev. B}\ }\textbf {\bibinfo {volume} {79}},\ \bibinfo
  {pages} {100502} (\bibinfo {year} {2009})}\BibitemShut {NoStop}%
\bibitem [{\citenamefont {Fink}\ \emph {et~al.}(2011)\citenamefont {Fink},
  \citenamefont {Soltwisch}, \citenamefont {Geck}, \citenamefont {Schierle},
  \citenamefont {Weschke},\ and\ \citenamefont
  {B\"uchner}}]{PhysRevB.83.092503}%
  \BibitemOpen
  \bibfield  {author} {\bibinfo {author} {\bibfnamefont {J.}~\bibnamefont
  {Fink}}, \bibinfo {author} {\bibfnamefont {V.}~\bibnamefont {Soltwisch}},
  \bibinfo {author} {\bibfnamefont {J.}~\bibnamefont {Geck}}, \bibinfo {author}
  {\bibfnamefont {E.}~\bibnamefont {Schierle}}, \bibinfo {author}
  {\bibfnamefont {E.}~\bibnamefont {Weschke}}, \ and\ \bibinfo {author}
  {\bibfnamefont {B.}~\bibnamefont {B\"uchner}},\ }\href {\doibase
  10.1103/PhysRevB.83.092503} {\bibfield  {journal} {\bibinfo  {journal} {Phys.
  Rev. B}\ }\textbf {\bibinfo {volume} {83}},\ \bibinfo {pages} {092503}
  (\bibinfo {year} {2011})}\BibitemShut {NoStop}%
\bibitem [{\citenamefont {Li}\ \emph {et~al.}(2007)\citenamefont {Li},
  \citenamefont {H\"ucker}, \citenamefont {Gu}, \citenamefont {Tsvelik},\ and\
  \citenamefont {Tranquada}}]{PhysRevLett.99.067001}%
  \BibitemOpen
  \bibfield  {author} {\bibinfo {author} {\bibfnamefont {Q.}~\bibnamefont
  {Li}}, \bibinfo {author} {\bibfnamefont {M.}~\bibnamefont {H\"ucker}},
  \bibinfo {author} {\bibfnamefont {G.~D.}\ \bibnamefont {Gu}}, \bibinfo
  {author} {\bibfnamefont {A.~M.}\ \bibnamefont {Tsvelik}}, \ and\ \bibinfo
  {author} {\bibfnamefont {J.~M.}\ \bibnamefont {Tranquada}},\ }\href {\doibase
  10.1103/PhysRevLett.99.067001} {\bibfield  {journal} {\bibinfo  {journal}
  {Phys. Rev. Lett.}\ }\textbf {\bibinfo {volume} {99}},\ \bibinfo {pages}
  {067001} (\bibinfo {year} {2007})}\BibitemShut {NoStop}%
\bibitem [{\citenamefont {Kojima}\ \emph {et~al.}()\citenamefont {Kojima},
  \citenamefont {Uchida}, \citenamefont {Fudamoto}, \citenamefont {Gat},
  \citenamefont {Larkin}, \citenamefont {Uemura},\ and\ \citenamefont
  {Luke}}]{KOJIMA}%
  \BibitemOpen
  \bibfield  {author} {\bibinfo {author} {\bibfnamefont {K.~M.}\ \bibnamefont
  {Kojima}}, \bibinfo {author} {\bibfnamefont {S.}~\bibnamefont {Uchida}},
  \bibinfo {author} {\bibfnamefont {Y.}~\bibnamefont {Fudamoto}}, \bibinfo
  {author} {\bibfnamefont {I.~M.}\ \bibnamefont {Gat}}, \bibinfo {author}
  {\bibfnamefont {M.~I.}\ \bibnamefont {Larkin}}, \bibinfo {author}
  {\bibfnamefont {Y.~J.}\ \bibnamefont {Uemura}}, \ and\ \bibinfo {author}
  {\bibfnamefont {G.~M.}\ \bibnamefont {Luke}},\ }\href
  {http://cat.inist.fr/?aModele=afficheN\&cpsidt=15370193} {\bibfield
  {journal} {\bibinfo  {journal} {Physica. B, Condensed matter}\ }\textbf
  {\bibinfo {volume} {326}},\ \bibinfo {pages} {316}}\BibitemShut {NoStop}%
\bibitem [{\citenamefont {Daou}\ \emph {et~al.}(2010)\citenamefont {Daou},
  \citenamefont {Chang}, \citenamefont {Leboeuf}, \citenamefont
  {Cyr-Choini\`{e}re}, \citenamefont {Lalibert\'{e}}, \citenamefont
  {Doiron-Leyraud}, \citenamefont {Ramshaw}, \citenamefont {Liang},
  \citenamefont {Bonn}, \citenamefont {Hardy},\ and\ \citenamefont
  {Taillefer}}]{Daou2010}%
  \BibitemOpen
  \bibfield  {author} {\bibinfo {author} {\bibfnamefont {R.}~\bibnamefont
  {Daou}}, \bibinfo {author} {\bibfnamefont {J.}~\bibnamefont {Chang}},
  \bibinfo {author} {\bibfnamefont {D.}~\bibnamefont {Leboeuf}}, \bibinfo
  {author} {\bibfnamefont {O.}~\bibnamefont {Cyr-Choini\`{e}re}}, \bibinfo
  {author} {\bibfnamefont {F.}~\bibnamefont {Lalibert\'{e}}}, \bibinfo {author}
  {\bibfnamefont {N.}~\bibnamefont {Doiron-Leyraud}}, \bibinfo {author}
  {\bibfnamefont {B.~J.}\ \bibnamefont {Ramshaw}}, \bibinfo {author}
  {\bibfnamefont {R.}~\bibnamefont {Liang}}, \bibinfo {author} {\bibfnamefont
  {D.~A.}\ \bibnamefont {Bonn}}, \bibinfo {author} {\bibfnamefont {W.~N.}\
  \bibnamefont {Hardy}}, \ and\ \bibinfo {author} {\bibfnamefont
  {L.}~\bibnamefont {Taillefer}},\ }\href {\doibase 10.1038/nature08716}
  {\bibfield  {journal} {\bibinfo  {journal} {Nature}\ }\textbf {\bibinfo
  {volume} {463}},\ \bibinfo {pages} {519} (\bibinfo {year}
  {2010})}\BibitemShut {NoStop}%
\bibitem [{\citenamefont {Lawler}\ \emph {et~al.}(2010)\citenamefont {Lawler},
  \citenamefont {Fujita}, \citenamefont {Lee}, \citenamefont {Schmidt},
  \citenamefont {Kohsaka}, \citenamefont {Kim}, \citenamefont {Eisaki},
  \citenamefont {Uchida}, \citenamefont {Davis}, \citenamefont {Sethna},\ and\
  \citenamefont {Kim}}]{Lawler2010}%
  \BibitemOpen
  \bibfield  {author} {\bibinfo {author} {\bibfnamefont {M.~J.}\ \bibnamefont
  {Lawler}}, \bibinfo {author} {\bibfnamefont {K.}~\bibnamefont {Fujita}},
  \bibinfo {author} {\bibfnamefont {J.}~\bibnamefont {Lee}}, \bibinfo {author}
  {\bibfnamefont {A.~R.}\ \bibnamefont {Schmidt}}, \bibinfo {author}
  {\bibfnamefont {Y.}~\bibnamefont {Kohsaka}}, \bibinfo {author} {\bibfnamefont
  {C.~K.}\ \bibnamefont {Kim}}, \bibinfo {author} {\bibfnamefont
  {H.}~\bibnamefont {Eisaki}}, \bibinfo {author} {\bibfnamefont
  {S.}~\bibnamefont {Uchida}}, \bibinfo {author} {\bibfnamefont {J.~C.}\
  \bibnamefont {Davis}}, \bibinfo {author} {\bibfnamefont {J.~P.}\ \bibnamefont
  {Sethna}}, \ and\ \bibinfo {author} {\bibfnamefont {E.-A.}\ \bibnamefont
  {Kim}},\ }\href {\doibase 10.1038/nature09169} {\bibfield  {journal}
  {\bibinfo  {journal} {Nature}\ }\textbf {\bibinfo {volume} {466}},\ \bibinfo
  {pages} {347} (\bibinfo {year} {2010})}\BibitemShut {NoStop}%
\bibitem [{\citenamefont {Ghiringhelli}\ \emph {et~al.}(2012)\citenamefont
  {Ghiringhelli}, \citenamefont {Le~Tacon}, \citenamefont {Minola},
  \citenamefont {Blanco-Canosa}, \citenamefont {Mazzoli}, \citenamefont
  {Brookes}, \citenamefont {De~Luca}, \citenamefont {Frano}, \citenamefont
  {Hawthorn}, \citenamefont {He}, \citenamefont {Loew}, \citenamefont {Sala},
  \citenamefont {Peets}, \citenamefont {Salluzzo}, \citenamefont {Schierle},
  \citenamefont {Sutarto}, \citenamefont {Sawatzky}, \citenamefont {Weschke},
  \citenamefont {Keimer},\ and\ \citenamefont
  {Braicovich}}]{Ghiringhelli17082012}%
  \BibitemOpen
  \bibfield  {author} {\bibinfo {author} {\bibfnamefont {G.}~\bibnamefont
  {Ghiringhelli}}, \bibinfo {author} {\bibfnamefont {M.}~\bibnamefont
  {Le~Tacon}}, \bibinfo {author} {\bibfnamefont {M.}~\bibnamefont {Minola}},
  \bibinfo {author} {\bibfnamefont {S.}~\bibnamefont {Blanco-Canosa}}, \bibinfo
  {author} {\bibfnamefont {C.}~\bibnamefont {Mazzoli}}, \bibinfo {author}
  {\bibfnamefont {N.~B.}\ \bibnamefont {Brookes}}, \bibinfo {author}
  {\bibfnamefont {G.~M.}\ \bibnamefont {De~Luca}}, \bibinfo {author}
  {\bibfnamefont {A.}~\bibnamefont {Frano}}, \bibinfo {author} {\bibfnamefont
  {D.~G.}\ \bibnamefont {Hawthorn}}, \bibinfo {author} {\bibfnamefont
  {F.}~\bibnamefont {He}}, \bibinfo {author} {\bibfnamefont {T.}~\bibnamefont
  {Loew}}, \bibinfo {author} {\bibfnamefont {M.~M.}\ \bibnamefont {Sala}},
  \bibinfo {author} {\bibfnamefont {D.~C.}\ \bibnamefont {Peets}}, \bibinfo
  {author} {\bibfnamefont {M.}~\bibnamefont {Salluzzo}}, \bibinfo {author}
  {\bibfnamefont {E.}~\bibnamefont {Schierle}}, \bibinfo {author}
  {\bibfnamefont {R.}~\bibnamefont {Sutarto}}, \bibinfo {author} {\bibfnamefont
  {G.~A.}\ \bibnamefont {Sawatzky}}, \bibinfo {author} {\bibfnamefont
  {E.}~\bibnamefont {Weschke}}, \bibinfo {author} {\bibfnamefont
  {B.}~\bibnamefont {Keimer}}, \ and\ \bibinfo {author} {\bibfnamefont
  {L.}~\bibnamefont {Braicovich}},\ }\href {\doibase 10.1126/science.1223532}
  {\bibfield  {journal} {\bibinfo  {journal} {Science}\ }\textbf {\bibinfo
  {volume} {337}},\ \bibinfo {pages} {821} (\bibinfo {year}
  {2012})}\BibitemShut {NoStop}%
\bibitem [{\citenamefont {Chang}\ \emph {et~al.}(2012)\citenamefont {Chang},
  \citenamefont {Blackburn}, \citenamefont {Holmes}, \citenamefont
  {Christensen}, \citenamefont {Larsen}, \citenamefont {Mesot}, \citenamefont
  {Liang}, \citenamefont {Bonn}, \citenamefont {Hardy}, \citenamefont
  {Watenphul}, \citenamefont {Zimmermann}, \citenamefont {Forgan},\ and\
  \citenamefont {Hayden}}]{Chang2012a}%
  \BibitemOpen
  \bibfield  {author} {\bibinfo {author} {\bibfnamefont {J.}~\bibnamefont
  {Chang}}, \bibinfo {author} {\bibfnamefont {E.}~\bibnamefont {Blackburn}},
  \bibinfo {author} {\bibfnamefont {a.~T.}\ \bibnamefont {Holmes}}, \bibinfo
  {author} {\bibfnamefont {N.~B.}\ \bibnamefont {Christensen}}, \bibinfo
  {author} {\bibfnamefont {J.}~\bibnamefont {Larsen}}, \bibinfo {author}
  {\bibfnamefont {J.}~\bibnamefont {Mesot}}, \bibinfo {author} {\bibfnamefont
  {R.}~\bibnamefont {Liang}}, \bibinfo {author} {\bibfnamefont {D.~a.}\
  \bibnamefont {Bonn}}, \bibinfo {author} {\bibfnamefont {W.~N.}\ \bibnamefont
  {Hardy}}, \bibinfo {author} {\bibfnamefont {a.}~\bibnamefont {Watenphul}},
  \bibinfo {author} {\bibfnamefont {M.~V.}\ \bibnamefont {Zimmermann}},
  \bibinfo {author} {\bibfnamefont {E.~M.}\ \bibnamefont {Forgan}}, \ and\
  \bibinfo {author} {\bibfnamefont {S.~M.}\ \bibnamefont {Hayden}},\ }\href
  {\doibase 10.1038/nphys2456} {\bibfield  {journal} {\bibinfo  {journal}
  {Nature Physics}\ }\textbf {\bibinfo {volume} {8}},\ \bibinfo {pages} {871}
  (\bibinfo {year} {2012})}\BibitemShut {NoStop}%
\bibitem [{\citenamefont {Wilkins}\ \emph {et~al.}(2011)\citenamefont
  {Wilkins}, \citenamefont {Dean}, \citenamefont {Fink}, \citenamefont
  {H\"ucker}, \citenamefont {Geck}, \citenamefont {Soltwisch}, \citenamefont
  {Schierle}, \citenamefont {Weschke}, \citenamefont {Gu}, \citenamefont
  {Uchida}, \citenamefont {Ichikawa}, \citenamefont {Tranquada},\ and\
  \citenamefont {Hill}}]{PhysRevB.84.195101}%
  \BibitemOpen
  \bibfield  {author} {\bibinfo {author} {\bibfnamefont {S.~B.}\ \bibnamefont
  {Wilkins}}, \bibinfo {author} {\bibfnamefont {M.~P.~M.}\ \bibnamefont
  {Dean}}, \bibinfo {author} {\bibfnamefont {J.}~\bibnamefont {Fink}}, \bibinfo
  {author} {\bibfnamefont {M.}~\bibnamefont {H\"ucker}}, \bibinfo {author}
  {\bibfnamefont {J.}~\bibnamefont {Geck}}, \bibinfo {author} {\bibfnamefont
  {V.}~\bibnamefont {Soltwisch}}, \bibinfo {author} {\bibfnamefont
  {E.}~\bibnamefont {Schierle}}, \bibinfo {author} {\bibfnamefont
  {E.}~\bibnamefont {Weschke}}, \bibinfo {author} {\bibfnamefont
  {G.}~\bibnamefont {Gu}}, \bibinfo {author} {\bibfnamefont {S.}~\bibnamefont
  {Uchida}}, \bibinfo {author} {\bibfnamefont {N.}~\bibnamefont {Ichikawa}},
  \bibinfo {author} {\bibfnamefont {J.~M.}\ \bibnamefont {Tranquada}}, \ and\
  \bibinfo {author} {\bibfnamefont {J.~P.}\ \bibnamefont {Hill}},\ }\href
  {\doibase 10.1103/PhysRevB.84.195101} {\bibfield  {journal} {\bibinfo
  {journal} {Phys. Rev. B}\ }\textbf {\bibinfo {volume} {84}},\ \bibinfo
  {pages} {195101} (\bibinfo {year} {2011})}\BibitemShut {NoStop}%
\bibitem [{\citenamefont {Achkar}\ \emph {et~al.}(2012)\citenamefont {Achkar},
  \citenamefont {Sutarto}, \citenamefont {Mao}, \citenamefont {He},
  \citenamefont {Frano}, \citenamefont {Blanco-Canosa}, \citenamefont {{Le
  Tacon}}, \citenamefont {Ghiringhelli}, \citenamefont {Braicovich},
  \citenamefont {Minola}, \citenamefont {{Moretti Sala}}, \citenamefont
  {Mazzoli}, \citenamefont {Liang}, \citenamefont {Bonn}, \citenamefont
  {Hardy}, \citenamefont {Keimer}, \citenamefont {Sawatzky},\ and\
  \citenamefont {Hawthorn}}]{Achkar2012}%
  \BibitemOpen
  \bibfield  {author} {\bibinfo {author} {\bibfnamefont {A.~J.}\ \bibnamefont
  {Achkar}}, \bibinfo {author} {\bibfnamefont {R.}~\bibnamefont {Sutarto}},
  \bibinfo {author} {\bibfnamefont {X.}~\bibnamefont {Mao}}, \bibinfo {author}
  {\bibfnamefont {F.}~\bibnamefont {He}}, \bibinfo {author} {\bibfnamefont
  {A.}~\bibnamefont {Frano}}, \bibinfo {author} {\bibfnamefont
  {S.}~\bibnamefont {Blanco-Canosa}}, \bibinfo {author} {\bibfnamefont
  {M.}~\bibnamefont {{Le Tacon}}}, \bibinfo {author} {\bibfnamefont
  {G.}~\bibnamefont {Ghiringhelli}}, \bibinfo {author} {\bibfnamefont
  {L.}~\bibnamefont {Braicovich}}, \bibinfo {author} {\bibfnamefont
  {M.}~\bibnamefont {Minola}}, \bibinfo {author} {\bibfnamefont
  {M.}~\bibnamefont {{Moretti Sala}}}, \bibinfo {author} {\bibfnamefont
  {C.}~\bibnamefont {Mazzoli}}, \bibinfo {author} {\bibfnamefont
  {R.}~\bibnamefont {Liang}}, \bibinfo {author} {\bibfnamefont {D.~A.}\
  \bibnamefont {Bonn}}, \bibinfo {author} {\bibfnamefont {W.~N.}\ \bibnamefont
  {Hardy}}, \bibinfo {author} {\bibfnamefont {B.}~\bibnamefont {Keimer}},
  \bibinfo {author} {\bibfnamefont {G.~A.}\ \bibnamefont {Sawatzky}}, \ and\
  \bibinfo {author} {\bibfnamefont {D.~G.}\ \bibnamefont {Hawthorn}},\ }\href
  {\doibase 10.1103/PhysRevLett.109.167001} {\bibfield  {journal} {\bibinfo
  {journal} {Physical Review Letters}\ }\textbf {\bibinfo {volume} {109}},\
  \bibinfo {pages} {167001} (\bibinfo {year} {2012})},\ \Eprint
  {http://arxiv.org/abs/1207.3667} {1207.3667} \BibitemShut {NoStop}%
\bibitem [{\citenamefont {Blackburn}\ \emph {et~al.}(2013)\citenamefont
  {Blackburn}, \citenamefont {Chang}, \citenamefont {H\"ucker}, \citenamefont
  {Holmes}, \citenamefont {Christensen}, \citenamefont {Liang}, \citenamefont
  {Bonn}, \citenamefont {Hardy}, \citenamefont {R\"utt}, \citenamefont
  {Gutowski}, \citenamefont {Zimmermann}, \citenamefont {Forgan},\ and\
  \citenamefont {Hayden}}]{PhysRevLett.110.137004}%
  \BibitemOpen
  \bibfield  {author} {\bibinfo {author} {\bibfnamefont {E.}~\bibnamefont
  {Blackburn}}, \bibinfo {author} {\bibfnamefont {J.}~\bibnamefont {Chang}},
  \bibinfo {author} {\bibfnamefont {M.}~\bibnamefont {H\"ucker}}, \bibinfo
  {author} {\bibfnamefont {A.~T.}\ \bibnamefont {Holmes}}, \bibinfo {author}
  {\bibfnamefont {N.~B.}\ \bibnamefont {Christensen}}, \bibinfo {author}
  {\bibfnamefont {R.}~\bibnamefont {Liang}}, \bibinfo {author} {\bibfnamefont
  {D.~A.}\ \bibnamefont {Bonn}}, \bibinfo {author} {\bibfnamefont {W.~N.}\
  \bibnamefont {Hardy}}, \bibinfo {author} {\bibfnamefont {U.}~\bibnamefont
  {R\"utt}}, \bibinfo {author} {\bibfnamefont {O.}~\bibnamefont {Gutowski}},
  \bibinfo {author} {\bibfnamefont {M.~v.}\ \bibnamefont {Zimmermann}},
  \bibinfo {author} {\bibfnamefont {E.~M.}\ \bibnamefont {Forgan}}, \ and\
  \bibinfo {author} {\bibfnamefont {S.~M.}\ \bibnamefont {Hayden}},\ }\href
  {\doibase 10.1103/PhysRevLett.110.137004} {\bibfield  {journal} {\bibinfo
  {journal} {Phys. Rev. Lett.}\ }\textbf {\bibinfo {volume} {110}},\ \bibinfo
  {pages} {137004} (\bibinfo {year} {2013})}\BibitemShut {NoStop}%
\bibitem [{\citenamefont {Blanco-Canosa}\ \emph {et~al.}(2013)\citenamefont
  {Blanco-Canosa}, \citenamefont {Frano}, \citenamefont {Loew}, \citenamefont
  {Lu}, \citenamefont {Porras}, \citenamefont {Ghiringhelli}, \citenamefont
  {Minola}, \citenamefont {Mazzoli}, \citenamefont {Braicovich}, \citenamefont
  {Schierle}, \citenamefont {Weschke}, \citenamefont {Le~Tacon},\ and\
  \citenamefont {Keimer}}]{PhysRevLett.110.187001}%
  \BibitemOpen
  \bibfield  {author} {\bibinfo {author} {\bibfnamefont {S.}~\bibnamefont
  {Blanco-Canosa}}, \bibinfo {author} {\bibfnamefont {A.}~\bibnamefont
  {Frano}}, \bibinfo {author} {\bibfnamefont {T.}~\bibnamefont {Loew}},
  \bibinfo {author} {\bibfnamefont {Y.}~\bibnamefont {Lu}}, \bibinfo {author}
  {\bibfnamefont {J.}~\bibnamefont {Porras}}, \bibinfo {author} {\bibfnamefont
  {G.}~\bibnamefont {Ghiringhelli}}, \bibinfo {author} {\bibfnamefont
  {M.}~\bibnamefont {Minola}}, \bibinfo {author} {\bibfnamefont
  {C.}~\bibnamefont {Mazzoli}}, \bibinfo {author} {\bibfnamefont
  {L.}~\bibnamefont {Braicovich}}, \bibinfo {author} {\bibfnamefont
  {E.}~\bibnamefont {Schierle}}, \bibinfo {author} {\bibfnamefont
  {E.}~\bibnamefont {Weschke}}, \bibinfo {author} {\bibfnamefont
  {M.}~\bibnamefont {Le~Tacon}}, \ and\ \bibinfo {author} {\bibfnamefont
  {B.}~\bibnamefont {Keimer}},\ }\href {\doibase
  10.1103/PhysRevLett.110.187001} {\bibfield  {journal} {\bibinfo  {journal}
  {Phys. Rev. Lett.}\ }\textbf {\bibinfo {volume} {110}},\ \bibinfo {pages}
  {187001} (\bibinfo {year} {2013})}\BibitemShut {NoStop}%
\bibitem [{\citenamefont {Dunsiger}\ \emph {et~al.}(2008)\citenamefont
  {Dunsiger}, \citenamefont {Zhao}, \citenamefont {Yamani}, \citenamefont
  {Buyers}, \citenamefont {Dabkowska},\ and\ \citenamefont
  {Gaulin}}]{PhysRevB.77.224410}%
  \BibitemOpen
  \bibfield  {author} {\bibinfo {author} {\bibfnamefont {S.~R.}\ \bibnamefont
  {Dunsiger}}, \bibinfo {author} {\bibfnamefont {Y.}~\bibnamefont {Zhao}},
  \bibinfo {author} {\bibfnamefont {Z.}~\bibnamefont {Yamani}}, \bibinfo
  {author} {\bibfnamefont {W.~J.~L.}\ \bibnamefont {Buyers}}, \bibinfo {author}
  {\bibfnamefont {H.~A.}\ \bibnamefont {Dabkowska}}, \ and\ \bibinfo {author}
  {\bibfnamefont {B.~D.}\ \bibnamefont {Gaulin}},\ }\href {\doibase
  10.1103/PhysRevB.77.224410} {\bibfield  {journal} {\bibinfo  {journal} {Phys.
  Rev. B}\ }\textbf {\bibinfo {volume} {77}},\ \bibinfo {pages} {224410}
  (\bibinfo {year} {2008})}\BibitemShut {NoStop}%
\bibitem [{\citenamefont {Abbamonte}(2006)}]{PhysRevB.74.195113}%
  \BibitemOpen
  \bibfield  {author} {\bibinfo {author} {\bibfnamefont {P.}~\bibnamefont
  {Abbamonte}},\ }\href {\doibase 10.1103/PhysRevB.74.195113} {\bibfield
  {journal} {\bibinfo  {journal} {Phys. Rev. B}\ }\textbf {\bibinfo {volume}
  {74}},\ \bibinfo {pages} {195113} (\bibinfo {year} {2006})}\BibitemShut
  {NoStop}%
\bibitem [{\citenamefont {Hinkov}\ \emph {et~al.}(2007)\citenamefont {Hinkov},
  \citenamefont {Bourges}, \citenamefont {Pailh\`{e}s}, \citenamefont {Sidis},
  \citenamefont {Ivanov}, \citenamefont {Frost}, \citenamefont {Perring},
  \citenamefont {Lin}, \citenamefont {Chen},\ and\ \citenamefont
  {Keimer}}]{Hinkov2007}%
  \BibitemOpen
  \bibfield  {author} {\bibinfo {author} {\bibfnamefont {V.}~\bibnamefont
  {Hinkov}}, \bibinfo {author} {\bibfnamefont {P.}~\bibnamefont {Bourges}},
  \bibinfo {author} {\bibfnamefont {S.}~\bibnamefont {Pailh\`{e}s}}, \bibinfo
  {author} {\bibfnamefont {Y.}~\bibnamefont {Sidis}}, \bibinfo {author}
  {\bibfnamefont {A.}~\bibnamefont {Ivanov}}, \bibinfo {author} {\bibfnamefont
  {C.~D.}\ \bibnamefont {Frost}}, \bibinfo {author} {\bibfnamefont {T.~G.}\
  \bibnamefont {Perring}}, \bibinfo {author} {\bibfnamefont {C.~T.}\
  \bibnamefont {Lin}}, \bibinfo {author} {\bibfnamefont {D.~P.}\ \bibnamefont
  {Chen}}, \ and\ \bibinfo {author} {\bibfnamefont {B.}~\bibnamefont
  {Keimer}},\ }\href {\doibase 10.1038/nphys720} {\bibfield  {journal}
  {\bibinfo  {journal} {Nature Physics}\ }\textbf {\bibinfo {volume} {3}},\
  \bibinfo {pages} {780} (\bibinfo {year} {2007})}\BibitemShut {NoStop}%
\bibitem [{\citenamefont {Liang}\ \emph {et~al.}(2006)\citenamefont {Liang},
  \citenamefont {Bonn},\ and\ \citenamefont {Hardy}}]{PhysRevB.73.180505}%
  \BibitemOpen
  \bibfield  {author} {\bibinfo {author} {\bibfnamefont {R.}~\bibnamefont
  {Liang}}, \bibinfo {author} {\bibfnamefont {D.~A.}\ \bibnamefont {Bonn}}, \
  and\ \bibinfo {author} {\bibfnamefont {W.~N.}\ \bibnamefont {Hardy}},\ }\href
  {\doibase 10.1103/PhysRevB.73.180505} {\bibfield  {journal} {\bibinfo
  {journal} {Phys. Rev. B}\ }\textbf {\bibinfo {volume} {73}},\ \bibinfo
  {pages} {180505} (\bibinfo {year} {2006})}\BibitemShut {NoStop}%
\bibitem [{\citenamefont {Hannon}\ \emph {et~al.}(1988)\citenamefont {Hannon},
  \citenamefont {Trammell}, \citenamefont {Blume},\ and\ \citenamefont
  {Gibbs}}]{PhysRevLett.61.1245}%
  \BibitemOpen
  \bibfield  {author} {\bibinfo {author} {\bibfnamefont {J.~P.}\ \bibnamefont
  {Hannon}}, \bibinfo {author} {\bibfnamefont {G.~T.}\ \bibnamefont
  {Trammell}}, \bibinfo {author} {\bibfnamefont {M.}~\bibnamefont {Blume}}, \
  and\ \bibinfo {author} {\bibfnamefont {D.}~\bibnamefont {Gibbs}},\ }\href
  {\doibase 10.1103/PhysRevLett.61.1245} {\bibfield  {journal} {\bibinfo
  {journal} {Phys. Rev. Lett.}\ }\textbf {\bibinfo {volume} {61}},\ \bibinfo
  {pages} {1245} (\bibinfo {year} {1988})}\BibitemShut {NoStop}%
\bibitem [{\citenamefont {Le~Tacon}\ \emph {et~al.}(2011)\citenamefont
  {Le~Tacon}, \citenamefont {Ghiringhelli}, \citenamefont {Chaloupka},
  \citenamefont {Sala}, \citenamefont {Hinkov}, \citenamefont {Haverkort},
  \citenamefont {Minola}, \citenamefont {Bakr}, \citenamefont {Zhou},
  \citenamefont {Blanco-Canosa}, \citenamefont {Monney}, \citenamefont {Song},
  \citenamefont {Sun}, \citenamefont {Lin}, \citenamefont {De~Luca},
  \citenamefont {Salluzzo}, \citenamefont {Khaliullin}, \citenamefont
  {Schmitt}, \citenamefont {Braicovich},\ and\ \citenamefont
  {Keimer}}]{NPh000294485400021}%
  \BibitemOpen
  \bibfield  {author} {\bibinfo {author} {\bibfnamefont {M.}~\bibnamefont
  {Le~Tacon}}, \bibinfo {author} {\bibfnamefont {G.}~\bibnamefont
  {Ghiringhelli}}, \bibinfo {author} {\bibfnamefont {J.}~\bibnamefont
  {Chaloupka}}, \bibinfo {author} {\bibfnamefont {M.~M.}\ \bibnamefont {Sala}},
  \bibinfo {author} {\bibfnamefont {V.}~\bibnamefont {Hinkov}}, \bibinfo
  {author} {\bibfnamefont {M.~W.}\ \bibnamefont {Haverkort}}, \bibinfo {author}
  {\bibfnamefont {M.}~\bibnamefont {Minola}}, \bibinfo {author} {\bibfnamefont
  {M.}~\bibnamefont {Bakr}}, \bibinfo {author} {\bibfnamefont {K.~J.}\
  \bibnamefont {Zhou}}, \bibinfo {author} {\bibfnamefont {S.}~\bibnamefont
  {Blanco-Canosa}}, \bibinfo {author} {\bibfnamefont {C.}~\bibnamefont
  {Monney}}, \bibinfo {author} {\bibfnamefont {Y.~T.}\ \bibnamefont {Song}},
  \bibinfo {author} {\bibfnamefont {G.~L.}\ \bibnamefont {Sun}}, \bibinfo
  {author} {\bibfnamefont {C.~T.}\ \bibnamefont {Lin}}, \bibinfo {author}
  {\bibfnamefont {G.~M.}\ \bibnamefont {De~Luca}}, \bibinfo {author}
  {\bibfnamefont {M.}~\bibnamefont {Salluzzo}}, \bibinfo {author}
  {\bibfnamefont {G.}~\bibnamefont {Khaliullin}}, \bibinfo {author}
  {\bibfnamefont {T.}~\bibnamefont {Schmitt}}, \bibinfo {author} {\bibfnamefont
  {L.}~\bibnamefont {Braicovich}}, \ and\ \bibinfo {author} {\bibfnamefont
  {B.}~\bibnamefont {Keimer}},\ }\href {\doibase 10.1038/NPHYS2041} {\bibfield
  {journal} {\bibinfo  {journal} {Nature Physics}\ }\textbf {\bibinfo {volume}
  {7}},\ \bibinfo {pages} {725} (\bibinfo {year} {2011})}\BibitemShut {NoStop}%
\bibitem [{\citenamefont {{Dean}}\ \emph
  {et~al.}(2013{\natexlab{a}})\citenamefont {{Dean}}, \citenamefont {{Dellea}},
  \citenamefont {{Minola}}, \citenamefont {{Wilkins}}, \citenamefont {{Konik}},
  \citenamefont {{Gu}}, \citenamefont {{Le Tacon}}, \citenamefont {{Brookes}},
  \citenamefont {{Yakhou-Harris}}, \citenamefont {{Kummer}}, \citenamefont
  {{Hill}}, \citenamefont {{Braicovich}},\ and\ \citenamefont
  {{Ghiringhelli}}}]{2013arXiv1305.1317D}%
  \BibitemOpen
  \bibfield  {author} {\bibinfo {author} {\bibfnamefont {M.~P.~M.}\
  \bibnamefont {{Dean}}}, \bibinfo {author} {\bibfnamefont {G.}~\bibnamefont
  {{Dellea}}}, \bibinfo {author} {\bibfnamefont {M.}~\bibnamefont {{Minola}}},
  \bibinfo {author} {\bibfnamefont {S.~B.}\ \bibnamefont {{Wilkins}}}, \bibinfo
  {author} {\bibfnamefont {R.~M.}\ \bibnamefont {{Konik}}}, \bibinfo {author}
  {\bibfnamefont {G.~D.}\ \bibnamefont {{Gu}}}, \bibinfo {author}
  {\bibfnamefont {M.}~\bibnamefont {{Le Tacon}}}, \bibinfo {author}
  {\bibfnamefont {N.~B.}\ \bibnamefont {{Brookes}}}, \bibinfo {author}
  {\bibfnamefont {F.}~\bibnamefont {{Yakhou-Harris}}}, \bibinfo {author}
  {\bibfnamefont {K.}~\bibnamefont {{Kummer}}}, \bibinfo {author}
  {\bibfnamefont {J.~P.}\ \bibnamefont {{Hill}}}, \bibinfo {author}
  {\bibfnamefont {L.}~\bibnamefont {{Braicovich}}}, \ and\ \bibinfo {author}
  {\bibfnamefont {G.}~\bibnamefont {{Ghiringhelli}}},\ }\href@noop {}
  {\bibfield  {journal} {\bibinfo  {journal} {ArXiv e-prints}\ } (\bibinfo
  {year} {2013}{\natexlab{a}})},\ \Eprint {http://arxiv.org/abs/1305.1317}
  {arXiv:1305.1317 [cond-mat.supr-con]} \BibitemShut {NoStop}%
\bibitem [{\citenamefont {Ghiringhelli}\ \emph {et~al.}(2004)\citenamefont
  {Ghiringhelli}, \citenamefont {Brookes}, \citenamefont {Annese},
  \citenamefont {Berger}, \citenamefont {Dallera}, \citenamefont {Grioni},
  \citenamefont {Perfetti}, \citenamefont {Tagliaferri},\ and\ \citenamefont
  {Braicovich}}]{PhysRevLett.92.117406}%
  \BibitemOpen
  \bibfield  {author} {\bibinfo {author} {\bibfnamefont {G.}~\bibnamefont
  {Ghiringhelli}}, \bibinfo {author} {\bibfnamefont {N.~B.}\ \bibnamefont
  {Brookes}}, \bibinfo {author} {\bibfnamefont {E.}~\bibnamefont {Annese}},
  \bibinfo {author} {\bibfnamefont {H.}~\bibnamefont {Berger}}, \bibinfo
  {author} {\bibfnamefont {C.}~\bibnamefont {Dallera}}, \bibinfo {author}
  {\bibfnamefont {M.}~\bibnamefont {Grioni}}, \bibinfo {author} {\bibfnamefont
  {L.}~\bibnamefont {Perfetti}}, \bibinfo {author} {\bibfnamefont
  {A.}~\bibnamefont {Tagliaferri}}, \ and\ \bibinfo {author} {\bibfnamefont
  {L.}~\bibnamefont {Braicovich}},\ }\href {\doibase
  10.1103/PhysRevLett.92.117406} {\bibfield  {journal} {\bibinfo  {journal}
  {Phys. Rev. Lett.}\ }\textbf {\bibinfo {volume} {92}},\ \bibinfo {pages}
  {117406} (\bibinfo {year} {2004})}\BibitemShut {NoStop}%
\bibitem [{\citenamefont {{Dean}}\ \emph
  {et~al.}(2013{\natexlab{b}})\citenamefont {{Dean}}, \citenamefont {{Dellea}},
  \citenamefont {{Springell}}, \citenamefont {{Yakhou-Harris}}, \citenamefont
  {{Kummer}}, \citenamefont {{Brookes}}, \citenamefont {{Liu}}, \citenamefont
  {{Sun}}, \citenamefont {{Strle}}, \citenamefont {{Schmitt}}, \citenamefont
  {{Braicovich}}, \citenamefont {{Ghiringhelli}}, \citenamefont {{Bozovic}},\
  and\ \citenamefont {{Hill}}}]{2013arXiv1303.5359D}%
  \BibitemOpen
  \bibfield  {author} {\bibinfo {author} {\bibfnamefont {M.~P.~M.}\
  \bibnamefont {{Dean}}}, \bibinfo {author} {\bibfnamefont {G.}~\bibnamefont
  {{Dellea}}}, \bibinfo {author} {\bibfnamefont {R.~S.}\ \bibnamefont
  {{Springell}}}, \bibinfo {author} {\bibfnamefont {F.}~\bibnamefont
  {{Yakhou-Harris}}}, \bibinfo {author} {\bibfnamefont {K.}~\bibnamefont
  {{Kummer}}}, \bibinfo {author} {\bibfnamefont {N.~B.}\ \bibnamefont
  {{Brookes}}}, \bibinfo {author} {\bibfnamefont {X.}~\bibnamefont {{Liu}}},
  \bibinfo {author} {\bibfnamefont {Y.-J.}\ \bibnamefont {{Sun}}}, \bibinfo
  {author} {\bibfnamefont {J.}~\bibnamefont {{Strle}}}, \bibinfo {author}
  {\bibfnamefont {T.}~\bibnamefont {{Schmitt}}}, \bibinfo {author}
  {\bibfnamefont {L.}~\bibnamefont {{Braicovich}}}, \bibinfo {author}
  {\bibfnamefont {G.}~\bibnamefont {{Ghiringhelli}}}, \bibinfo {author}
  {\bibfnamefont {I.}~\bibnamefont {{Bozovic}}}, \ and\ \bibinfo {author}
  {\bibfnamefont {J.~P.}\ \bibnamefont {{Hill}}},\ }\href@noop {} {\bibfield
  {journal} {\bibinfo  {journal} {ArXiv e-prints}\ } (\bibinfo {year}
  {2013}{\natexlab{b}})},\ \Eprint {http://arxiv.org/abs/1303.5359}
  {arXiv:1303.5359 [cond-mat.supr-con]} \BibitemShut {NoStop}%
\bibitem [{\citenamefont {Coneri}\ \emph {et~al.}(2010)\citenamefont {Coneri},
  \citenamefont {Sanna}, \citenamefont {Zheng}, \citenamefont {Lord},\ and\
  \citenamefont {De~Renzi}}]{PhysRevB.81.104507}%
  \BibitemOpen
  \bibfield  {author} {\bibinfo {author} {\bibfnamefont {F.}~\bibnamefont
  {Coneri}}, \bibinfo {author} {\bibfnamefont {S.}~\bibnamefont {Sanna}},
  \bibinfo {author} {\bibfnamefont {K.}~\bibnamefont {Zheng}}, \bibinfo
  {author} {\bibfnamefont {J.}~\bibnamefont {Lord}}, \ and\ \bibinfo {author}
  {\bibfnamefont {R.}~\bibnamefont {De~Renzi}},\ }\href {\doibase
  10.1103/PhysRevB.81.104507} {\bibfield  {journal} {\bibinfo  {journal} {Phys.
  Rev. B}\ }\textbf {\bibinfo {volume} {81}},\ \bibinfo {pages} {104507}
  (\bibinfo {year} {2010})}\BibitemShut {NoStop}%
\bibitem [{\citenamefont {Rigamonti}\ \emph {et~al.}(1998)\citenamefont
  {Rigamonti}, \citenamefont {Borsa},\ and\ \citenamefont
  {Carretta}}]{0034-4885-61-10-002}%
  \BibitemOpen
  \bibfield  {author} {\bibinfo {author} {\bibfnamefont {A.}~\bibnamefont
  {Rigamonti}}, \bibinfo {author} {\bibfnamefont {F.}~\bibnamefont {Borsa}}, \
  and\ \bibinfo {author} {\bibfnamefont {P.}~\bibnamefont {Carretta}},\ }\href
  {http://stacks.iop.org/0034-4885/61/i=10/a=002} {\bibfield  {journal}
  {\bibinfo  {journal} {Reports on Progress in Physics}\ }\textbf {\bibinfo
  {volume} {61}},\ \bibinfo {pages} {1367} (\bibinfo {year}
  {1998})}\BibitemShut {NoStop}%
\bibitem [{\citenamefont {Haug}\ \emph {et~al.}(2010)\citenamefont {Haug},
  \citenamefont {Hinkov}, \citenamefont {Sidis}, \citenamefont {Bourges},
  \citenamefont {Christensen}, \citenamefont {Ivanov}, \citenamefont {Keller},
  \citenamefont {Lin},\ and\ \citenamefont {Keimer}}]{1367-2630-12-10-105006}%
  \BibitemOpen
  \bibfield  {author} {\bibinfo {author} {\bibfnamefont {D.}~\bibnamefont
  {Haug}}, \bibinfo {author} {\bibfnamefont {V.}~\bibnamefont {Hinkov}},
  \bibinfo {author} {\bibfnamefont {Y.}~\bibnamefont {Sidis}}, \bibinfo
  {author} {\bibfnamefont {P.}~\bibnamefont {Bourges}}, \bibinfo {author}
  {\bibfnamefont {N.~B.}\ \bibnamefont {Christensen}}, \bibinfo {author}
  {\bibfnamefont {A.}~\bibnamefont {Ivanov}}, \bibinfo {author} {\bibfnamefont
  {T.}~\bibnamefont {Keller}}, \bibinfo {author} {\bibfnamefont {C.~T.}\
  \bibnamefont {Lin}}, \ and\ \bibinfo {author} {\bibfnamefont
  {B.}~\bibnamefont {Keimer}},\ }\href
  {http://stacks.iop.org/1367-2630/12/i=10/a=105006} {\bibfield  {journal}
  {\bibinfo  {journal} {New Journal of Physics}\ }\textbf {\bibinfo {volume}
  {12}},\ \bibinfo {pages} {105006} (\bibinfo {year} {2010})}\BibitemShut
  {NoStop}%
\end{thebibliography}%

\end{document}